\documentstyle[epsfig,prl,twocolumn,aps]{revtex}
\begin{document}
\twocolumn[\hsize\textwidth\columnwidth\hsize\csname @twocolumnfalse\endcsname
\bibliographystyle{prsty} 
\title{Solving the Poisson-Boltzmann Equation to Obtain Interaction
Energies Between Confined, Like-charged Cylinders}
\author{Mark Ospeck \and Seth Fraden \\ The Complex Fluids Group,
Martin Fisher School of Physics, Brandeis University, \\Waltham, MA
~02454, USA}
\date{\today} \maketitle
\begin{abstract}
We numerically solve the non-linear Poisson-Boltzmann
equation for two cylinders confined by two parallel charged
plates. The repulsive electrical double layer
component of the cylinder pair potential is
substantially reduced by confinement between
like-charged plates.  While the effective cylinder
surface charge is increased by
the confinement, the effective interaction screening length is reduced,
this effect being dominant so that the repulsive confined cylinder-cylinder
interaction potential is reduced.
\end{abstract}
\vskip2pc]

\bf I. INTRODUCTION \rm

Recent experiments have cast doubts that the DLVO pair potential is
correctly describing the pair interaction between like-charged colloids
in aqueous suspension in a confined region where the colloid motions are
being restricted by the confining double layers. Long range attractive
potentials of order 1 $\rm K_{B}T$ in strength have been observed
\cite{KF,CG,CG2,GNat,C-TC-RA-L}. Additionally there is interest in
whether shorter range interactions between like-charged cylinders in
monovalent electrolyte can become attractive under certain circumstances
\cite{B,PRP,TWTJ,G-JMBG}. Here  we examine the problem of long range
interactions between parallel like-charged cylinders confined between
like-charged plates by numerically solving the two dimensional,
confined, non-linear Poisson-Boltzmann equation.

\bf II. NUMERICALLY SOLVING THE 2D, \\
CONFINED,NON-LINEAR POISSON-BOLTZMANN \\ EQUATION \rm

Two water-solvated like-charged cylinders experience an electrical
double layer (EDL) repulsion, their behavior governed by the non-linear
Poisson-Boltzmann (PB) equation, which dictates both the potential and
simple ion concentration distributions in their vicinity:
\begin{equation}
\nabla^2 \phi = \kappa^2 \sinh\phi
\end{equation}
Where $\nabla^2$ is the scalar Laplacian operating upon the
dimensionless potential $\phi$, which equals the scalar potential $\psi$
divided by $\rm K_{B}T/ \it ze$, with $e$ being the quantum of charge,
and $z$ the charge of a single counterion. $\kappa^{-1}$ is the Debye
screening length. Assuming the simple ions to be monovalent $\rm K_{B}T/
\it ze$= 25.69mV at 298 K.

The EDL interaction forms the repulsive component of the DLVO potential
between two like-charged colloidal particles.  At constant thermodynamic
volume the Helmholtz free energy is appropriate for describing the EDL
interaction between two like-charged colloidal particles. At constant
surface potential, the EDL interaction energy has three parts: an
attractive electrical term  (negative cylinder surface residues
attracted to the positive charge counterion cloud between them), a
repulsive entropy term (ion/solvent configurational entropy decreasing
with decreasing cylinder separation), and a repulsive chemical potential
term (counterion number is decreased by surface charge condensation as
the cylinders approach each other). While at constant cylinder surface
charge, the EDL has only two parts: an attractive electrical term and a
dominant repulsive entropy term (with constant surface charge,
counterion number is constant).

\begin{figure}[t] \noindent \centering
\epsfig{file=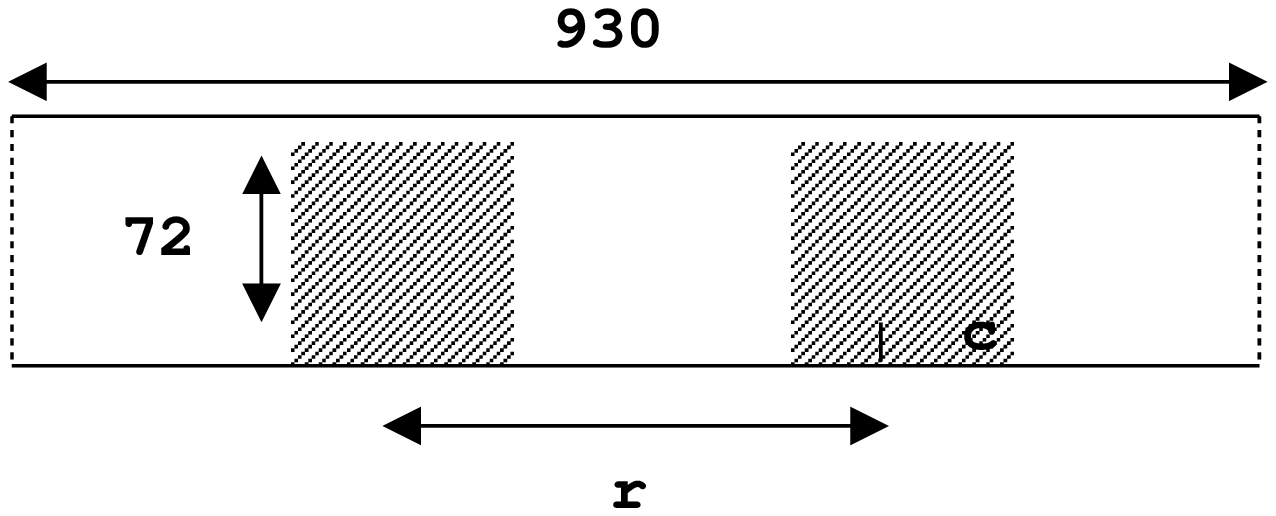,width=8.2cm}
\renewcommand{\baselinestretch}{0.9}\small\normalsize
\caption{\label{Fig1}
\protect\begin{small}
Geometry for solving the 2d, confined, non-linear Poisson-Boltzmann
equation. Lengths are in grids of a 2d square lattice, where the
reservoir, or bulk screening length is 6-30 grids. Typically, a constant
potential boundary condition (b.c.) is imposed on the confining line
charges, a constant charge (potential slope) b.c. is imposed on the
circles, and a free b.c. is employed at the left and right boundaries of
the space (extrapolating the local potential slope, which equals zero
far from the circles). The space's width was chosen so that the
circle-circle interaction potentials were insensitive to increases in
width. $c$ is the closest distance between the circle and the line, and
$r$ is the center-center separation between the circles, all in grids.
\protect\end{small}
}
\end{figure}

The two dimensional problem of circles confined by line charges is
equivalent to parallel cylinders confined by walls in three dimensions.
Numerical computations of the nonlinear Poisson-Boltzmann (PB) equation
were performed principally for two circles of constant (dimensionless)
charge density ($\sigma= \frac {d\phi_{\rm circ}}{d \hat n}$ where $\hat
n$ is the normal to the surface) confined by two constant
(dimensionless) potential line charges ($\phi_{\rm line}$) as sketched
in Fig.~\ref{Fig1}. The boundary condition was usually constant charge
on the cylinders and always constant potential on the confining charged
walls because we supposed the cylinders to possess strong acid groups,
typical of surface groups such as polystyrene sulfonate, while we
assumed the confining walls were made of glass, which contains a high
density of weak acid silanol groups.

Our use of a constant potential glass boundary condition needs further
clarification. This is a standard low surface potential, weak acid
surface group boundary condition \cite{Bergna} and is enforced at the
outer Helmholz plane (OHP) where the compact, or Stern layer ends, and
by association dissociation equilibria of weak acid surface groups and
also two dimensional mobility of counterions within the compact layer.
This layer is a highly concentrated monolayer of aqueous counterions
typically containing  90 percent of the glass countercharge and well
over 100 millivolts of potential change so that the resulting glass
boundary condition which faces the electrolyte at the OHP is a low
(usually well less than 100 millivolt) constant potential
\cite{Bergna,VO}.  Our results depend upon this glass regulating
constant potential boundary condition's ability to stand up under
compression. Should it fail then there would follow electric field lines
leaking into the low dielectric glass and polarizing electron clouds. It
would then become important to account for unscreened interactions
between these polarized electron clouds. This paper discusses the
limited constant potential confining boundary case. A complete treatment
would include the field inside the glass, model the glass-water
interface with a regularizing boundary condition \cite{CPW}, as well as
calculating the field between the charged circles. Although it is
expected that the constant potential glass boundary condition would
remain valid at compressions between surfaces down to one screening
length \cite{I} there is some evidence that under extreme compression it
is in fact the case that field lines are traversing through the external
dielectric \cite{CG2}. The inclusion of regularizing boundary conditions
\cite{CPW} is a direction for future research.

There is zero electric field within the circles. Note that the constant
surface charge boundary condition on the circles will not be a problem
because field line penetration into the low dielectric circles causes
only a small perturbation in the circle-circle interaction (see Fig.9).

A periodic boundary condition was employed at the left and right edges
of the space (at these edges we extrapolated the local potential slope
in the direction parallel to the confining lines, which was equal to
zero, since these boundaries were many tens of screening lengths
separated from the circle edges; see Fig.4). The space's width was
chosen so that the circle-circle interaction potentials were insensitive
to increases in this width.

The geometrical details of our numerical simulation are as follows. Our
circle's radius $a$ was fixed at 36 grids of a 2d square lattice, the
screening length $\kappa^{-1}$ varied from 6 to 30 grids, while the
space was made 930 grids in length in order to avoid edge effects. The
distance from the circle's surface to the line charge $c$ was varied
from 6 to $\frac{1}{2}$ screening lengths.  Venturing much below $\kappa
c= \frac{1}{2}$ was found to introduce coarse graining errors, and a
more sophisticated multi-grid relaxation algorithm \cite {NumRecC} would
be necessary in order to accurately track the highly curved potential
function in this very strongly confined regime. Potential fields for
fixed circle and line boundary conditions, $c$, and $\kappa^{-1}$ were
obtained by numerically solving the non-linear Poisson-Boltzmann
equation on a Silicon Graphics work station having a MIPS R4400
processor, and by using a successive overrelaxation (SOR) \cite
{NumRecC} algorithm, employing an SOR factor of 1.85. Total free
energies $TFE$ for a given separation $r$ were obtained via the
Helmholtz prescription from Overbeek \cite{O}, so that a circle-circle
interaction potential could subsequently be constructed.  Then we made a
two parameter fit of the Helmholtz interaction energy to
\begin{equation}
V_{EDL,2D}= \frac{Z^2}{L^2} \; GF \frac {e^{-\kappa_{local}r}}
{\sqrt{r}},
\end{equation}
$\frac{Z}{L}$ being the cylinder's charge per unit length. The fit
netted $\kappa^{-1}_{local}$ and a geometric prefactor $GF$ depending
upon local screening length and particle radius. Between spheres in
three dimensions \cite{Sood}
\begin{equation}
GF_{3D} = \frac{ e^{\kappa_{local}2a}}{(1+ \kappa_{local} a)^2},
\end{equation}
while between cylinders in two dimensions
\begin{equation}
GF_{2D} = \frac{ e^{\kappa_{local}2a}}{(1+ 2\kappa_{local} a)^2}.
\end{equation}
These fits revealed that use of the bulk screening length $\kappa^{-1}$
was not appropriate for the confined circle-circle interaction.  As the
circles were increasingly confined by the line charges the circle-circle
interaction's effective screening length was found to decrease in a
systematic way, with the line charge counterions screening the
circle-circle interaction.  In addition, the circle's effective charge $
\frac{Z}{L} \; \sqrt{GF}$ was found to be increased by the confinement.

\begin{figure}[t] \noindent \centering
\epsfig{file=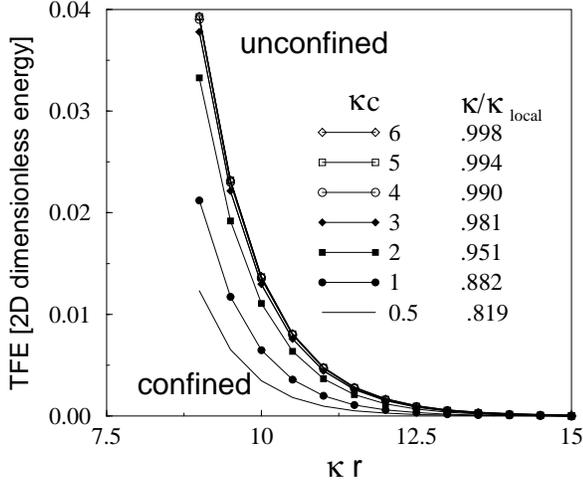,width=8.2cm}
\renewcommand{\baselinestretch}{0.9}\small\normalsize
\caption{\label{Fig2}
\protect\begin{small}
Reduction of interparticle repulsion between a pair of charged rods due
to the rods being confined by a pair of charged plates. 2d dimensionless
total free energies $TFE$ versus circle-circle separations $\kappa r$
for the case where inverse screening length times circle radius $\kappa
a$=1.8, and as a function of circle edge-line separation $\kappa c$.
Note that in order to obtain a three dimensional free energy one must
multiply $TFE$ by $\frac{\rm K_B T}{4 \pi}$ and by the number of Bjerrum
lengths the circular cylinders extend into the $z$ direction. The circle
boundary condition (b.c.) is .0507 constant potential slope
(corresponding to a -1 free surface potential if $\kappa a$=1.8) and the
confining line b.c. is -2 constant potential.  Lines are two parameter
fits of the data points to Eq. 2, thereby obtaining the geometric factor
$GF$ and $\kappa^{-1}_{local}$ as a function of circle edge-line
separation $\kappa c$. The repulsive interaction is substantially
reduced and the local Debye screening length is shortened by the
presence of the charged plates.
\protect\end{small}
}
\end{figure}

Overbeek's Free Energy bookkeeping \cite{O} goes as follows:
\begin{equation}
\rm Total \: Free \: Energy \it =TFE=EFE+CFE
\end{equation}
\begin{equation}
EFE=\rm Electrical \: Free \: Energy \it =EE-T\Delta S
\end{equation}
\begin{eqnarray}
CFE^* & = & \rm Chemical \: Free \: Energy^* \it =  \nonumber \\
      & = & -2\int_{S}(\phi \frac{d\phi}{d \hat n })dS \nonumber \\ & =
      & -2\int_A ((\nabla\phi)^2 +\kappa^2\phi \sinh\phi)dA
\end{eqnarray}
\begin{equation}
EE^*=\rm Electrical \: Energy^* \it = \int_A (\nabla\phi)^2 dA
\end{equation}
\begin{equation}
T \Delta S^*=-2\kappa^2\int_A (1-\cosh\phi+\phi \sinh\phi)dA
\end{equation}
Where * indicates a dimensionless 2d energy, $dS$ represents a small
element of constant potential boundary, and $dA$ a small area of
electrolyte. Converting a 2d dimensionless free energy into a 3d
dimensionfull one requires multiplying by $\rm k_{B}T$, by $\frac{1}{4
\pi}$, and by one half the number of Bjerrum lengths the circular
cylinders extend into the z direction \cite{SP}.  Take Fig.~\ref{Fig2}
as an example: if its cylinders were 500nm in radius,
$\kappa^{-1}$=280nm (in order that $\kappa a$=1.8), and the cylinders
were 1000nm in length, then we should multiply their dimensionless 2d
energies by  $\rm k_{B}T
\frac{1}{4 \pi} \frac{1000 \rm nm}{1.4 \rm nm}$
= 57 $\rm k_{B}T$, i.e. at $\kappa r$=10, $\kappa c$=6  the cylinders would
experience about a 0.6 $\rm k_{B}T$ repulsion. The free energies
obtained from numerical solutions of the PB for interacting flat plates
compared well against the tabulated values in the Verwey-Overbeek
monograph \cite{VO}, and also against Israelachvili's figure 12.10
\cite{I}.

Figure 2 shows seven EDL potential barriers between two circles confined
by line charges arranged in the geometry depicted in Fig.4. The
dimensionless total free energy $TFE$ (Eqs.5-9) is plotted versus the
circle's center to center separation $r$. The circles have a free
potential equal to -1, but it is their surface charges which are held
constant during the circle-circle interaction, this because the circle
surfaces are assumed to posess strong acid groups which resist charge
condensation.  Their constant potential slopes are equal to .0507, the
appropriate slope for a -1 potential free surface when $\kappa a$=1.8,
i.e. the screening length is 20 grids, and the circle's radius $a$ is
fixed at 36 grids. Here the confining line charges are held at constant
-2. potential, and are thus said to be perfectly regulating because
their surfaces are assumed to be composed out of weak acid groups.
Circle edge-line charge separation $c$ is varied from $\frac{1}{2}$ to 6
$\kappa^{-1}$.  Circle motion is assumed to be adiabatically cut off
from the motion of the ion gas, i.e. the ions are assumed to readjust to
the new circle configurations with extreme rapidity. Circle separation
$r$ begins at 15$\kappa^{-1}$ and is decremented by units of .5
$\kappa^{-1}$ down to 9 $\kappa^{-1}$, all the while solving the PB
equation and subsequently recording the configuration's Helmholtz free
energy $TFE$.  A two parameter fit of the ($r$,$TFE$) data points is
made to the 2d EDL potential function (Eq.2), thereby obtaining the
interaction screening length $\kappa^{-1}_{local}$, and observing that
this circle-circle interaction screening length is decreased from the
bulk value by close confinement of the line charges.

Figure 3 breaks down the relative contributions made to the
circle-circle interaction potential between the electrical, entropic,
and chemical free energies (Eqs.5-9) for Fig.2's case of strongly
confined circles. The case of two constant potential line charges
confining two constant charge circles contains two attractive and one
repulsive terms. The smaller of the attractive terms is the chemical
free energy term (the number of \it line-charge \rm counterions
increasing as the circles move together). The larger attractive term is
electrical in origin arising from the counterions located in between the
two circles---like in a hydrogen molecule. However, at 300 Kelvin, the
repulsive entropic term dominates the attractive terms.

\begin{figure}[t] \noindent \centering
\epsfig{file=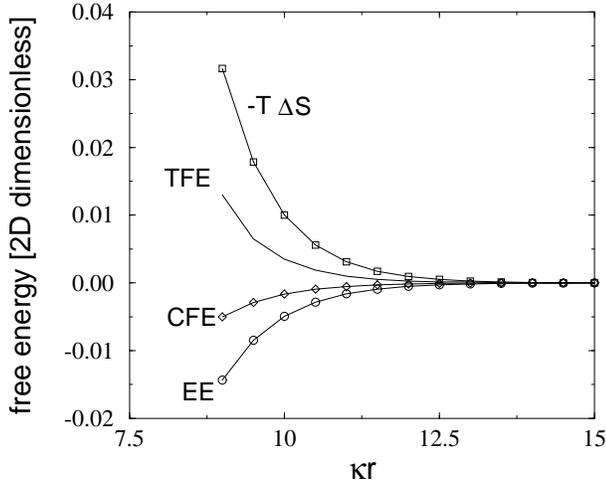,width=8.2cm}
\renewcommand{\baselinestretch}{0.9}\small\normalsize
\caption{\label{Fig3}
\protect\begin{small}
The three component terms of the circle-circle interaction free energy
for Fig.2's case of strong confinement are plotted versus circle
center-center separation $\kappa r$, where $EE$ is electrical energy,
always attractive, $-T\Delta S$ due to ion/solvent configurational
entropy, always repulsive, and $CFE$ is a weak attractive chemical free
energy (line charge counterion number increases when circles are close
together).  Notice that the repulsive ion/solvent configurational
entropy term is dominant, meaning that the total potential is repulsive
and that this repulsion is due to osmotic pressure forces. The
conditions are two constant -2 potential line charges strongly confining
($\kappa c$=.5) two interacting constant potential slope circles (.0507
slope appropriate to a free -1 potential if $\kappa a=1.8$).
\protect\end{small}
}
\end{figure}

Figure 4 shows three contour plots associated with Fig.2 and appropriate
to greater or lesser degrees of confinement.  When the circles are more
confined the contours surrounding each circle become more elongated,
reflecting the fact that many of the circle's electric field lines are
now terminating locally in the line charge's double layer.  One could
think as if the field lines were terminating in partial image charges
created in the line charge double layer.

\begin{figure}[t] \noindent \centering
\epsfig{file=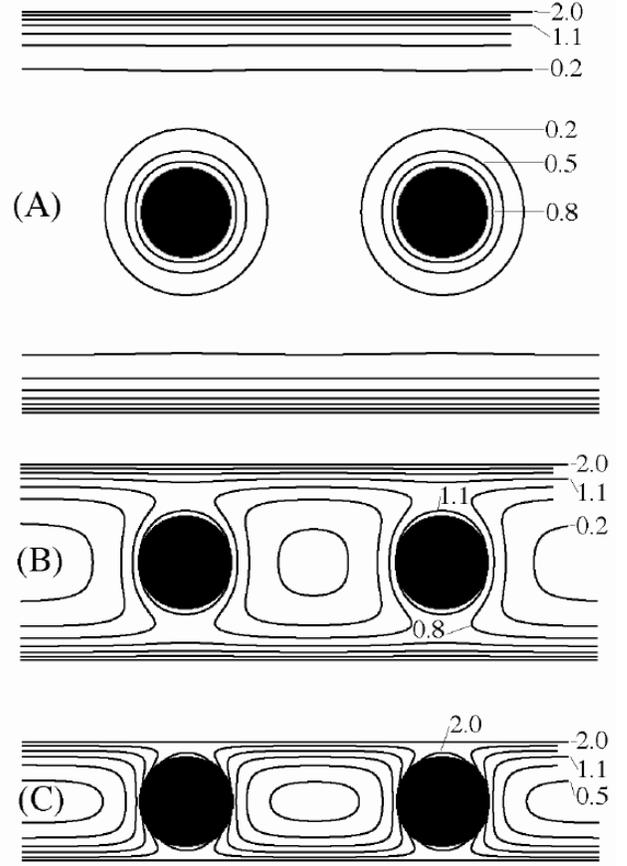,width=8.2cm}
\renewcommand{\baselinestretch}{0.9}\small\normalsize
\caption{\label{Fig4}
\protect\begin{small}
Contour plots of the dimensionless potential ($\phi$) for the
conditions shown in Fig.2. The circles have constant charge (their
boundary potential slope which equals
.0507 corresponds to an unconfined potential equal to -1 when $\kappa a$=1.8)
and the line charges are at constant -2 potential. All potentials are
negative. $\kappa r$=10 for all three plots. There is no electric field
within the black circle interiors. (a) The circle-circle interaction is
virtually undisturbed by the distant charged lines, $\kappa c$=6. (b)
The circle-circle interaction is becoming weakened by confinement,
$\kappa c$=2. Note the increasing circle surface potential. (c) Strong
confinement, $\kappa c$=.5. Field lines normally involved in the
circle-circle repulsive interaction are being redirected, terminating on
the counterions in the charged line's double layer.
\protect\end{small}
}
\end{figure}

Figures 5,6, and 7 are all connected. In Fig.5 the confining line
potential is increased to -6 and the confined circle-circle interaction
screening length $\kappa^{-1}_{local}$ is appreciably lessened when
compared with Fig.2 which has only a -2 potential confining line. The
large line capacitance is acting to hide the circles from each other.
Figure 6 gives the relationship between confining line potential and
local screening length, while Fig.7 shows for a free
\it line charge \rm the relationship between confining
line potential $ \phi_{line}$ and line capacitance per unit length
$C/L$.

\begin{equation}
 C/L = \frac{(\frac {d\phi_{\rm line \it}} {d \hat n})}{\phi_{\rm line}}
\end{equation}

\begin{figure}[t] \noindent \centering
\epsfig{file=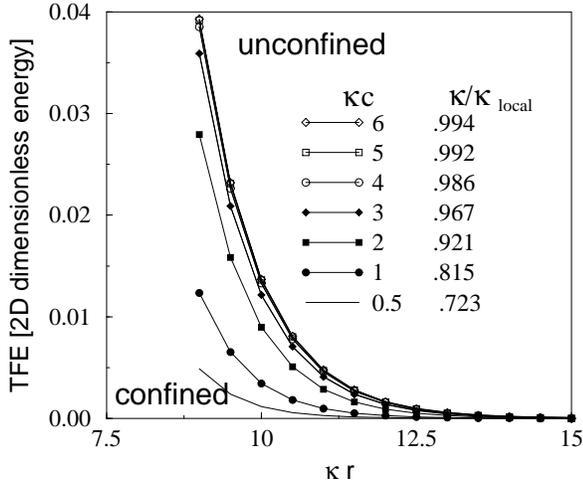,width=8.2cm}
\renewcommand{\baselinestretch}{0.9}\small\normalsize
\caption{\label{Fig5}
\protect\begin{small}
Reduction of interparticle repulsion by increasing the confining
potentials compared to Fig.2. Plots of total free energy $TFE$ versus
circle-circle separation ($\kappa r$) for
.0507 constant potential slope circles
confined by line charges held at a -6 potential, and $\kappa a$=1.8. The
circle-circle barrier potential is reduced by approximately one order of
magnitude when the circles are severely confined. Note that the degree
of confinement $\kappa c$ is varied from 6 to $\frac{1}{2}$.
\protect\end{small}
}
\end{figure}
\begin{figure}[t] \noindent \centering
\epsfig{file=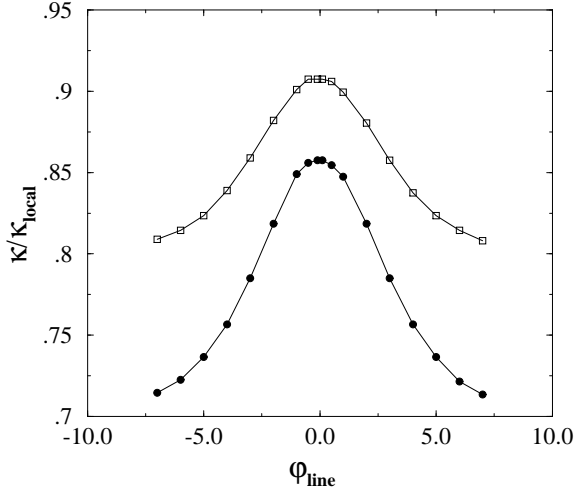,width=8.2cm}
\renewcommand{\baselinestretch}{0.9}\small\normalsize
\caption{\label{Fig6}
\protect\begin{small}
The ratio of local screening length $\kappa^{-1}_{local}$ to bulk
screening length $\kappa^{-1}$ as a function of the confining line
potential $\phi_{line}$ boundary condition. The curve is symmetric with
respect to the confining line potential. The conditions are $\kappa
a$=1.8, solid circles correspond to $ \kappa c=.5$ (strong confinement),
hollow squares to $\kappa c =1$ (weaker confinement), while the circles
have constant (charge) potential slope =.0507.
\protect\end{small}
}
\end{figure}

Figure 7's potential slope fits the Grahame equation \cite{I}
\begin{equation}
\frac{\rm d \phi}{\rm d \it \hat n}= \kappa 2 \sinh{\frac{\phi}{2}}.
\end{equation}
Also note Fig.7 capacitance per unit length's nonzero y-intercept
($\approx .05=\kappa$) shows that line capacitance goes like inverse
screening length $\kappa$ for a free line when $\phi_{line}$ is held at
zero potential (obtained by dividing the r.h.s. of the Grahame equation
by $\phi_{line}$, while taking the limit as $\phi_{\rm line} \rightarrow
0$). In Fig.6 we plotted local screening length versus confining line
potential and obtained gaussian functions symmetric about the zero of
potential. Notice that the local screening length is .85$\kappa^{-1}$
for zero potential line charges strongly confining the circle-circle
interaction, $\kappa c$=.5, and that this is connected with the fact
that a zero potential confining line charge posesses an increased
capacitance determined by a Grahame equation modified to account for the
confinement \cite{CPW}. Essentially, strong confinement brings the zero
potential line an additional capacitance due to its sharing of the
confined circle's counterions. Interestingly, the local interaction
screening length is independent of the sign of the surface charge on the
confining lines. Apparently as regards the local screening length it
does not matter if a negative field line sourced from a circle
terminates in the line's double layer on a positive counterion or on the
positive line charge itself.

\begin{figure}[t] \noindent \centering
\epsfig{file=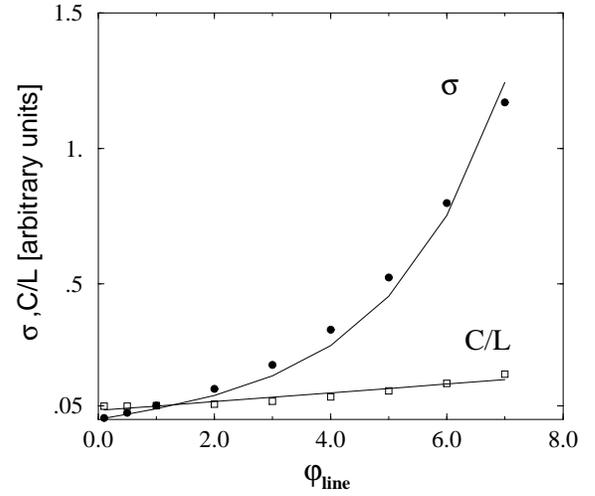,width=8.2cm}
\renewcommand{\baselinestretch}{0.9}\small\normalsize
\caption{\label{Fig7}
\protect\begin{small}
Plots for a free \it line charge \rm of its boundary potential
slope $\sigma$($=\frac {d \phi_{\rm line}}{d \hat n}$, solid circles)
and its intrinsic capacitance per unit length $C/L$ ($=
\frac{\sigma}{\phi_{\rm line}}$, hollow squares; see Eq.10) versus free line
potential $\phi_{line}$. The line tracing the solid circles is a fit to
the Grahame equation (Eq.11).
\protect\end{small}
}
\end{figure}

The second Eq.2 fitting parameter $GF$ stands for the DLVO \it geometric
factor \rm \cite{Sood}, which accounts for the internal volume of the
cylinder being excluded to the screening ions. For a typical colloidal
particle there are two effects competing to set effective surface charge
$Z \sqrt{GF}$ as one increases the concentration of potential
determining ions (p.d.i.). A p.d.i. electrolyte is contrasted to an
indifferent electrolyte, the distinction being that an indifferent
electrolyte is unable to bind to surface residues.  The first effect
concerns binding of a p.d.i. to a surface residue, thereby decreasing
$Z$, while the second occurs as the concentration gradient of p.d.i.'s
in the vicinity of a charged surface is increased, thus making the
particle appear to have increased its surface charge. This second effect
is the one accounted for by DLVO's $GF$.  If one holds $Z$ constant
while increasing ionic strength, then one observes an increasing
effective particle surface charge $Z^{\star}= Z \: \sqrt{GF}$ due to the
steepened counterion concentration gradient. We see in Fig.8 that $GF$
increases in a gaussian fashion with increasingly negative confining
line potential. Compare the behaviors of the two fitting parameters $GF$
(Fig.8) and $\kappa^{-1}_{local}$ (Fig.6).  Local screening length
doesn't care about the sign of the surrounding line charge, but $GF$
certainly does.

\begin{figure}[t] \noindent \centering
\epsfig{file=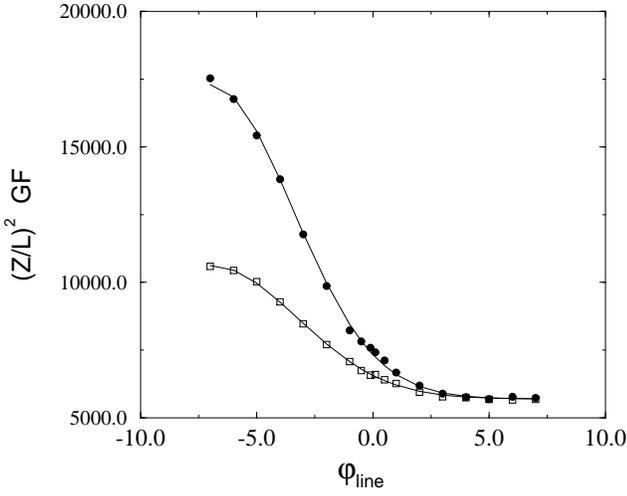,width=8.2cm}
\renewcommand{\baselinestretch}{0.9}\small\normalsize
\caption{\label{Fig8}
\protect\begin{small}
Plot of $\frac {Z^2}{L^2} \; GF = \frac {Z^{\star 2}}{L^2}$
versus line potential $\phi_{line}$ for two confined interacting
.0507 constant potential slope circles when $\kappa a$=1.8.
$\frac{Z^{\star}}{L}$ is the effective surface charge on the circle. The
solid circular data points correspond to $ \kappa c =.5$ (strong
confinement), and the hollow squares to $ \kappa c =1$ (weaker
confinement).
\protect\end{small}
}
\end{figure}

After $\kappa^{-1}_{local}$ as a function of $c$ is obtained from Eq.2
(see Figs.2 and 5) we make a one parameter fit to the function
\begin{equation}
\kappa^{-1}_{local}= \kappa^{-1} e^{-Ae^{-\kappa c}}
\end{equation}
and plot the result in Fig.9. Local screening length goes like the
inverse square root of local ionic strength ($\kappa^{-1}_{local} \sim
n^{-.5}$), which itself goes like an exponential function of the local
potential (Boltzmann factor: $n \sim e^{- \phi}$), and finally this
local potential goes like a decaying exponential function of distance
from a charged surface ($\phi \sim e^{- \kappa c}$).

\begin{figure}[t] \noindent \centering
\epsfig{file=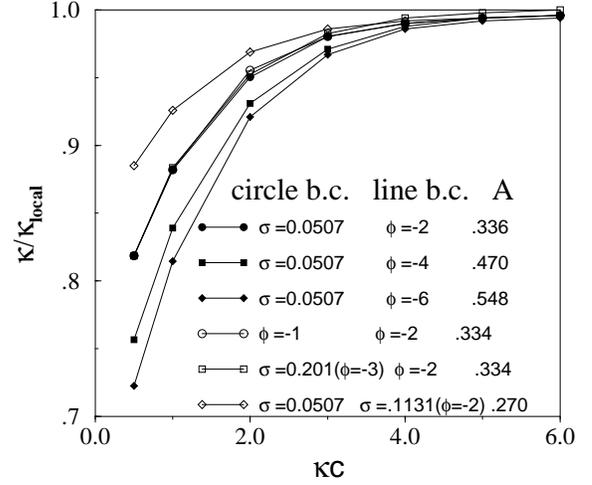,width=8.2cm}
\renewcommand{\baselinestretch}{0.9}\small\normalsize
\caption{\label{Fig9}
\protect\begin{small}
Local screening length versus the confinement distance between
the circle and the line charge, $\kappa c$, for various circle and line
boundary conditions, where $\phi=\frac {\psi}{K_BT/e}$ is a
dimensionless surface potential boundary condition (b.c.) and
$\sigma=\frac {d \phi}{d \hat n}$ a surface charge b.c., and with
$\kappa a=1.8$.  The lines are one parameter fits to the factor $A$ in
Eq.12. The solid data points demonstrate the line potential's control
over the local screening length. Meanwhile changing the circle b.c. had
little effect on interaction screening length, as seen by comparing the
solid circle, open circle, and open square data sets. Note that a
constant charge b.c. on the confining plate corresponding to a -2
potential, i.e. the open diamond ($A$=.270) had a significantly smaller
effect on local screening length than did its -2 confining potential
counterpart. This probably had to do with the fact that \it line charge
\rm counterion number increased when the circles were close together,
for the case of constant potential confining lines (see Fig.3).
\protect\end{small}
}
\end{figure}

Fig.10 shows that decreasing $c$, i.e. increasing the degree of
confinement, increases the $GF$ for all of the constant-charge
circle-circle interactions.  In contrast, for the case of confined
constant-potential circles the square data points show that the
combination $\frac{Z^2}{L^2} GF$  actually decreases.  For strong
confinement (small $c$) condensation of counterions decreases the
circle's surface charge $\frac{Z}{L}$, and this effect wins out over an
increasing $GF$.

\begin{figure}[t] \noindent \centering
\epsfig{file=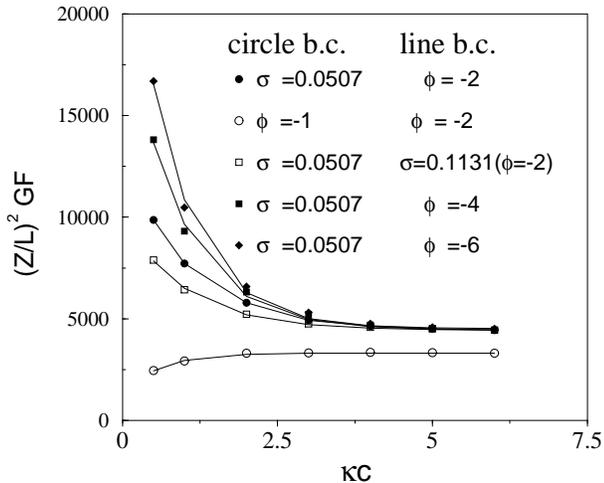,width=8.2cm}
\renewcommand{\baselinestretch}{0.9}\small\normalsize
\caption{\label{Fig10}
\protect\begin{small}
Squared surface charge per unit length times the DLVO \it geometric
factor \rm $\frac {Z^2}{L^2} \; GF$ (see Eq.2) versus confinement
$\kappa c$ for various circle and line boundary conditions, where
$\phi=\frac {\psi}{K_BT/e}$ is a dimensionless surface potential b.c.,
$\sigma =\frac {d \phi}{d \hat n} $ a surface charge b.c., and $\kappa
a=1.8$. Only in the constant potential circle case where surface charge
regulation acts to decrease $Z$ does $\frac{Z^2}{L^2} \; GF$ decrease
with increasing confinement.
\protect\end{small}
}
\end{figure}

Finally we vary bulk screening length from 6 grids to 30 grids for
various amounts of confinement, and obtain the dependence of local
screening length on bulk screening length for these differing degrees of
confinement (Fig.11).  In our 2d geometry there are 3 independent length
scales : circle radius (fixed at 36 grids), circle edge-line separation
$c$ (varying from 10 to 40 grids), and $\kappa^{-1}$ (varying from 6 to
30 grids, i.e. $\kappa a$ decreases from 6 to 1.2). The circle-circle
interaction's length scale $\kappa^{-1}_{local}$ is a function of these
three.  As one increases the bulk screening length one decreases the
space's midplane potential towards the line potential ($\phi_{line}=-2$
in Fig.11), and thus more of the circle's field lines traverse a region
where the local ionic strength has been increased by a factor of
$\cosh[-2]= 3.76$ times the bulk ionic strength (counterions increased
by the factor $e^2$ while the coions decrease by $e^{-2}$).  Hence the
screening length ``seen'' by these field lines will be reduced by a
factor $\frac{1}{\sqrt{3.76}}=.52$.  While for the case $\kappa^{-1}= 6$
most of the interaction field lines sample electrolyte where
$\kappa^{-1}_{local}=\kappa^{-1}$.

\begin{figure}[t] \noindent \centering
\epsfig{file=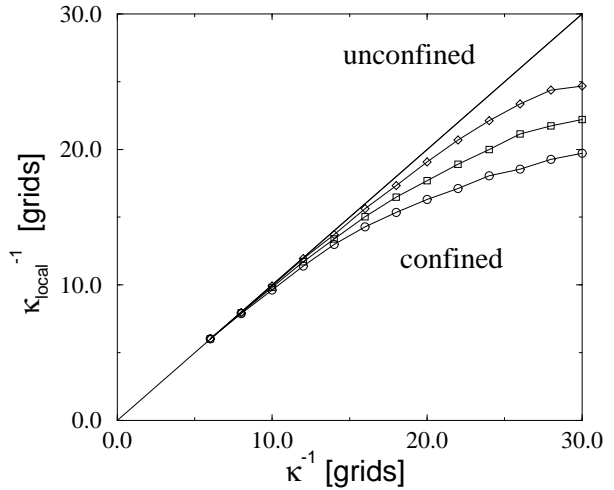,width=8.2cm}
\renewcommand{\baselinestretch}{0.9}\small\normalsize
\caption{\label{Fig11}
\protect\begin{small}
Local screening length $\kappa^{-1}_{local}$ as a function of bulk
screening length for various degrees of confinement.  The circles have a
constant potential slope =.0507 (corresponding to a -1 free surface
potential only when $\kappa a=1.8$) and the line charges are held at a
-2 potential.  Local screening lengths were obtained by a fit to Eq.2 of
the total free energies in a region where circle surface to surface
separation was approximately 5.5 bulk screening lengths. The diagonal
corresponds to $\kappa^{-1}_{local}$= $\kappa^{-1}$, open circles to
$c$=10 grids, open squares to $c$=20 grids, open diamonds to $c$=40
grids, while the circle radius is 36 grids. The local screening length
increases to the bulk screening length as the confining line charges
move apart.
\protect\end{small}
}
\end{figure}

Solutions to the confined non-linear Poisson-Boltzmann equation for
like- charged cylinders appear to be repulsive, screened, Yukawa-like
potentials, having an effective screening length $\kappa^{-1}_{local}$
which is found to be a decreasing function of increasing confinement.
The effective charge of the cylinders $Z^{\star}=\frac{Z}{L} \sqrt{GF}$
is found to be an increasing function of increasing confinement.

\bf III. CONCLUSIONS \rm

By numerically solving the two dimensional, non-linear, confined
Poisson-Boltmann equation we've found that the electrical double layer
(EDL) repulsion can be significantly decreased between two parallel
cylinders when confined between two charged plates because the confining
double layers constitute a high capacitance region which screens the
cylinder-cylinder interaction.

\bf ACKNOWLEDGMENTS \rm
We thank the referee for numerous suggestions for clarifying the text.
This research was supported by DOE DE-FG02-87ER45084.

\end{document}